\documentclass[prd,reprint,nofootinbib,showpacs,superscriptaddress]{revtex4-1}
\usepackage{graphicx} 
\usepackage{hyperref}
\usepackage{amsfonts}
\usepackage{amsmath,amssymb}
\usepackage{bm} 
\usepackage{color}
\usepackage{epstopdf} 
\usepackage{epsfig}
\usepackage{subfig} 
\usepackage{float}

{\rm }

\def\be{\begin{equation}}
 \def\ee{\end{equation}}
 \def\bea{\begin{eqnarray}}
 \def\eea{\end{eqnarray}}
 \def\bes{\begin{eqnarray}}
 \def\ees{\end{eqnarray}}
 \def\bi{\begin{itemize}}
 \def\ei{\end{itemize}} 

\def\2{\frac{1}{2}}
\def\4{\frac{1}{4}}

\begin{document}

\title{True randomness from an incoherent source}

\author{Bing Qi}
\email{qib1@ornl.gov}
\affiliation{Quantum Information Science Group, Computational Sciences and Engineering Division,
Oak Ridge National Laboratory, Oak Ridge, TN 37831-6085, USA}
\affiliation{Department of Physics and Astronomy, The
University of Tennessee, Knoxville, TN 37996-1200, USA
}

\date{\today}

\begin{abstract}

Quantum random number generators (QRNGs) harness the intrinsic randomness in measurement processes: the measurement outputs are truly random given the input state is a superposition of the eigenstates of the measurement operators. In the case of \emph{trusted} devices, true randomness could be generated from a mixed state $\rho$ so long as the system entangled with $\rho$ is well protected. We propose a random number generation scheme based on measuring the quadrature fluctuations of a single mode thermal state using an optical homodyne detector. By mixing the output of a broadband amplified spontaneous emission (ASE) source with a single mode local oscillator (LO) at a beam splitter and performing differential photo-detection, we can selectively detect the quadrature fluctuation of a single mode output of the ASE source, thanks to the filtering function of the LO. Experimentally, a quadrature variance about three orders of magnitude larger than the vacuum noise has been observed, suggesting this scheme can tolerate much higher detector noise in comparison with QRNGs based on measuring the vacuum noise. The high quality of this entropy source is evidenced by the small correlation coefficients of the acquired data. A Toeplitz hashing extractor is applied to generate unbiased random bits from the Gaussian distributed raw data, achieving an efficiency of 5.12 bits per sample. The output of the Toeplitz extractor successfully passes all the NIST statistical tests for random numbers. \footnote{This manuscript has been authored by UT-Battelle, LLC under Contract No. DE-AC05-00OR22725 with the U.S. Department of Energy. The United States Government retains and the publisher, by accepting the article for publication, acknowledges that the United States Government retains a non-exclusive, paid-up, irrevocable, world-wide license to publish or reproduce the published form of this manuscript, or allow others to do so, for United States Government purposes. The Department of Energy will provide public access to these results of federally sponsored research in accordance with the DOE Public Access Plan (http://energy.gov/downloads/doe-public-access-plan).
}

\end{abstract}

\maketitle

Truly random numbers are required in many branches of science and technology, from fundamental research in
quantum mechanics \cite{H15} to practical applications such as cryptography \cite{S96}. While a pseudorandom number generator can expand a short random seed into a long train of apparent ``random'' bits using deterministic algorithms, the entropy of generated random numbers is still bounded by the original short random seed. To generate true randomness, researchers have been exploring various physical processes.

Quantum random number generation is an emerging technology \cite{MY16, HG17}, which can provide high-quality random numbers with proven randomness. Different from physical random number generators exploring chaotic behaviors of classical systems, a quantum random number generator (QRNG) harnesses the truly probabilistic nature of fundamental quantum processes \cite{JA00, SG00}.

In general, the process of random number generation can be divided into two steps: the measurement step and the randomness extraction step. In the first step, a well-defined measurement is performed on a well-characterized entropy source. Ideally, in this step, the measurement system should only detect the intrinsic quantum noise of the entropy source. In practice, both the source and the detection system are not perfect and will introduce additional technical noises. In the worst case of scenario, the technical noises could be accessible to (or even controlled by) a malicious adversary (Eve) and thus cannot be trusted. Furthermore, the raw output of the detector may not be uniformly distributed. The second step in random number generation is to perform randomness extraction to generate uniformly distributed random numbers uncorrelated to the untrusted technical noises \cite{MX13, FR13, LP14, HA15, XQ12, AA14, MA15, NH15}. In practice, to conduct randomness extraction effectively, the quantum noise should be dominant over the technical noises.

Among various QRNG implementations, schemes based on photonic technology have drawn a lot of attention for high rates, low cost and the potential of chip-size integration \cite{KE15, AA16}. Both single photon detectors and optical homodyne detectors have been employed in photonic QRNGs. The latter is especially appealing in practice since highly efficient photo-diodes working at room temperature can be applied. Several QRNG schemes based on optical homodyne detection, exploring fundamental noises such as vacuum fluctuation \cite{TV07, GW10, ST10, SA11} and laser phase noise \cite{QC10, GT10, JC11, YL14, ZV17}, have been studied extensively. Remarkably, QRNG based on laser phase noise has been employed in a recent loop-hole free Bell experiment \cite{AA15}.

Nevertheless, there are still practical challenges in existing schemes. In QRNG based on vacuum noise \cite{TV07, GW10, ST10, SA11}, one major source of technical noises is the electrical noise of the homodyne detector. We remark that building a high-speed homodyne detector with electrical noise significantly below the shot noise is technically challenging \cite{OH08, CQ11, KB12}. This may in turn limit the operating speed of this type of QRNG. In QRNG based on laser phase noise \cite{QC10, GT10, JC11, YL14}, fiber interferometers with large arm imbalance (on the order of nanoseconds) are commonly employed. To achieve high random number generation rates, either phase stabilization of the fiber interferometer \cite{QC10} or high speed modulation of the laser source \cite{JC11} is required. In a more recent chip-size design \cite{AA16}, instead of using a cumbersome fiber interferometer, the outputs from two independent distributed feedback (DFB) lasers are mixed at a beam splitter. Random numbers are generated by operating one laser in gain switching (GS) mode, while the other in continuous wave (CW) mode. Essentially, the laser in GS mode provides a train of phase randomized laser pulses, while the laser in CW mode acts as a phase reference in coherent detection. To achieve a high interference visibility, sophisticated temperature control is required to match and stabilize the wavelengths of the two lasers.

In this paper, we demonstrate a random number generation scheme based on measuring quadrature 
fluctuations of a single mode thermal state using an optical homodyne detector. This scheme is implemented by beating a broadband amplified spontaneous emission (ASE) source with a single mode local oscillator (LO) at a symmetric beam splitter and performing differential photo-detection. Similar to the design in \cite{AA16}, our scheme does not require a fiber interferometer of large arm imbalance, which makes it very appealing in chip-size integration. Furthermore, both the ASE source and the LO are operated in CW mode, no active intensity modulation or phase (and polarization) control is required. 

Note even though the output of the ASE source is broadband, the optical homodyne detector selectively detects photons in the same spatial-temporal (and polarization) mode as the LO. The intrinsic ``filtering'' function of the LO allows us to perform single mode measurement without actually preparing a single mode thermal state, which greatly simplifies the implementation. Since the bandwidth of the ASE source is tens of nanometer, it is very easy to align the central wavelength of the LO within the spectral range of the ASE source.

Our scheme is different from previous studies in \cite{WS10, LC11}, where direct detection (rather than coherent detection) is employed to measure the ASE-ASE beat noises. Comparing with QRNG based on measuring the vacuum noise, our scheme can tolerate much higher detector noises. This is because the quadrature variance of a single mode thermal state with an average photon number of n is 2n+1 times as large as that of vacuum noise \cite{L00}. By preparing a thermal state with a large average photon number, we can effectively increase the quantum noise-to-detector noise ratio.

At the first sight, it seems controversial to generate true randomness from an incoherent source since intrinsic randomness is deeply connected to quantum coherence \cite{YZ15}. We remark that while truly quantum randomness can only originate from a superposition (pure) state, the input state $\rho$ to the measurement device does not have to be pure so long as the system entangled with $\rho$ cannot be accessed by Eve. One illustrative example is the QRNG based on radioactivity, where electrons from a radioactive source, such as $^{90}$Sr, are registered by a Geiger-Mueller tube at random times \cite{II56, S70}. In this case, at the time right before the measurement, the joint state of the radioactive nuclei and the electron can be described by
\bes\label{eq1} \vert\psi\rangle=\alpha\vert U\rangle_n\vert 0\rangle_e+\beta\vert D\rangle_n\vert 1\rangle_e, \ees
where $\vert U\rangle_n$ ($\vert D\rangle_n$) represents undecayed (decayed) nuclei, $\vert 0\rangle_e$ ($\vert 1\rangle_e$) represents 0 (1) electron emitted, and $\alpha$ and $\beta$ are normalization coefficients.

While the state $\vert \psi \rangle$ in (1) is pure, the state accessible to the Geiger-Mueller tube (the measurement device) is a mixed state given by $\rho_e=\alpha^2\vert 0\rangle\langle 0\vert+\beta^2\vert 1\rangle\langle 1\vert$. To generate secure random numbers, one underlying assumption is Eve cannot access (or control) the radioactive source (otherwise, Eve may acquire a copy of the random bit by measuring the quantum state of the source). The same argument can also be applied to QRNGs based on laser phase noise \cite{QC10, GT10, JC11, YL14} or amplified spontaneous emission \cite{WS10, LC11}, where the emitted photons accessible to the detector are entangled with atoms in the light source. In this paper, our discussion is based on this ``trusted'' device scenario. Note this is different from the assumptions adopted in the so-called source-independent QRNG \cite{FS07, CZ16}, where the state input to the detector could be entangled with the environment accessible to Eve, thus cannot be trusted.

The experimental setup is shown in Fig.1. A fiber amplifier (PriTel, Inc.) with vacuum state input is employed as an incoherent broadband source. Previous studies have shown that the ASE noise generated by a fiber amplifier is thermal \cite{WH98, VV00}. We have conducted conjugate homodyne detection and verified the photon statistics of a single mode component (selected by the LO) of the ASE source follows Bose-Einstein distribution \cite{QL17}, as expected from a single mode thermal state. To facilitate the estimation of the photon number per mode arrived at the optical homodyne detector and reduce the power of unused light, a 0.8nm optical bandpass filter centered at 1542 nm is placed after the ASE source (BP in Fig.1). A laser source with a central wavelength of 1542 nm (Clarity-NLL-1542-HP from Wavelength Reference) is employed as the LO. Note it is not necessary to stabilize the laser frequency, since it can never drift out the above 0.8nm range under normal operation. To prevent the saturation of the photodetector, an optical attenuator is applied to reduce the LO power to about 4mW.
 
\begin{figure}[t]
	\includegraphics[width=.45\textwidth]{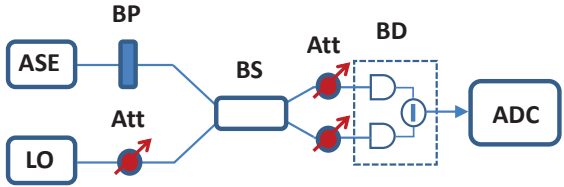}
	\captionsetup{justification=raggedright,
					singlelinecheck=false }
	\caption{Experimental setup. ASE-amplified spontaneous emission source; LO-local oscillator; BP-optical bandpass filter; Att-optical attenuator; BS-fiber beam splitter; BD-balanced photo-receiver;
ADC-analog-to-digital converter.} 
	\label{fig:1}
\end{figure}
 
The outputs from the ASE source and the LO laser beat at a 50:50 fiber beam splitter (BS in Fig.1) and the
differential interference signal is measured by a 350MHz balanced photo-receiver (from Thorlabs, BD in Fig.1). While the output of the ASE source is unpolarized, the LO will automatically pick out the same polarization mode as itself. No polarization control is required. The overall efficiency $\eta$ of the detection system, including coupling losses and insertion losses of optical components and the quantum efficiency of photodiode, has been determined to be 0.5. An analog-to-digital converter (ADC) is used to sample the output of the balanced receiver. All the optical components are single mode fiber pigtailed. 
 
The system shown in Fig.1 allows us to measure a randomly chosen (due to random phase relation between the
LO laser and the ASE source) quadrature of a single mode thermal state. The output is expected to be truly
random following a Gaussian distribution with a zero mean and a variance of 2n+1 in the shot-noise unit \cite{L00}, where n is the average photon number of the ASE source in the mode determined by the LO. To calibrate the average photon number n, an optical power meter is used to measure the output power P of the ASE source after the optical filter. Using the relation of $\nu=c/\lambda$, we can determine the total mode number N corresponding to a bandwidth of $\Delta\lambda$ and a time window of $\Delta t=1s$ as 
\bes\label{eq2} N=2\Delta\nu\Delta t=\frac{2c\Delta\lambda}{\lambda^2},\ees 
where $\lambda$ ($\nu$) is the central wavelength (frequency) of the filtered ASE light and c is the speed of light in vacuum. The factor 2 is due to the two polarization modes generated by the ASE source. The effective average photon number n (after taking into account of the detection efficiency of the optical homodyne detector) can be determined from the optical power P using
\bes\label{eq3} n=\frac{\eta P}{Nh\nu}=\frac{\eta P\lambda^3}{2hc^2\Delta\lambda},\ees 
where h is the Planck constant. Experimentally, P is measured to be 29.0$\mu W$. Given $\eta$=0.5, $\Delta\lambda$=0.8nm, and $\lambda$=1542nm, the average photon number n has been determined to be about 555 per mode. This suggests that the expected quadrature variance of this source is about three orders of magnitude larger than the vacuum noise, which is a significant advantage of this scheme.

In the first experiment, a 12-bit data acquisition board (Texas Instruments) was employed to sample the output of the homodyne detector. The maximum sampling rate is mainly limited by the bandwidth of the photodetector: if the sampling rate is close to or above the detector bandwidth, adjacent samples will show significant correlation. Given the detector bandwidth is 350MHz, a sampling rate of 100MHz was chosen in our experiments. Limited by the memory size of the data acquisition board, $10^5$ samples were collected in this experiment. The histogram of the raw data together with a Gaussian fit curve are shown in Fig.2. The raw data fits a Gaussian distribution reasonably well, as expected from a single mode thermal state. The deviation from a perfect Gaussian distribution can be attributed to the systematic errors of the data acquisition board, including its nonlinearity. 

\begin{figure}[t]
	\includegraphics[width=.5\textwidth]{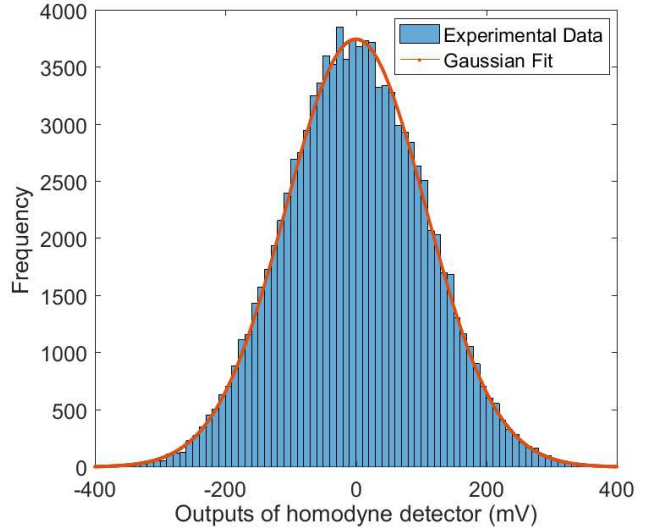}
	\captionsetup{justification=raggedright,
					singlelinecheck=false }
	\caption{Histogram of the measurement results and a Gaussian fit curve. Sample size=$10^5$.}
	\label{fig:2}
\end{figure}

To further justify the Gaussian assumption, we perform a chi-square test for goodness of fit \cite{PT92} using 1000 samples from the above raw data.\footnote{Note if the sample size is too large, the chi-square test for goodness of fit becomes extremely sensitive to even very small deviations from the ideal distribution. In practice, such a deviation is unavoidable due to systematic errors in the measurement system, such as the nonlinearity of the detector and the data acquisition board. As shown later, the random number generation rate is determined by the min-entropy of the experimental data. So a small deviation from the ideal Gaussian distribution will not compromise the quality of the generated random numbers.} The chi-square test yields an $\alpha$ value of 0.35, indicating that there is no good reason to reject the Gaussian hypothesis.

To determine the variance of the experimental data in the shot-noise unit, we also measured the detector noise (by turning off both the ASE source and the LO) and the vacuum noise (by turning on the LO only) separately. The measurement results are shown in Fig.3. Comparing Fig.2 with Fig.3, it is obvious that the quantum noise is much larger than both the vacuum noise and the detector noise. This allows us to apply low-cost, noisy detector to implement our scheme. By comparing the data in Fig.3 (a) with that in Fig.3 (b), the detector noise has been determined to be 0.62 in the shot noise limit. After being normalized to the vacuum noise, the variance of the Gaussian random numbers shown in Fig.2 has been determined to be 963 in the shot-noise unit, corresponding to a single mode thermal state with an average photon number of 481. Note this average photon number is about $13\%$ smaller than the value estimated from optical power P using (3). We suspect this discrepancy is mainly due to the errors in determining the
bandwidth of the optical filter and the efficiency of the detector.

\begin{figure}[t]
	\includegraphics[width=.5\textwidth]{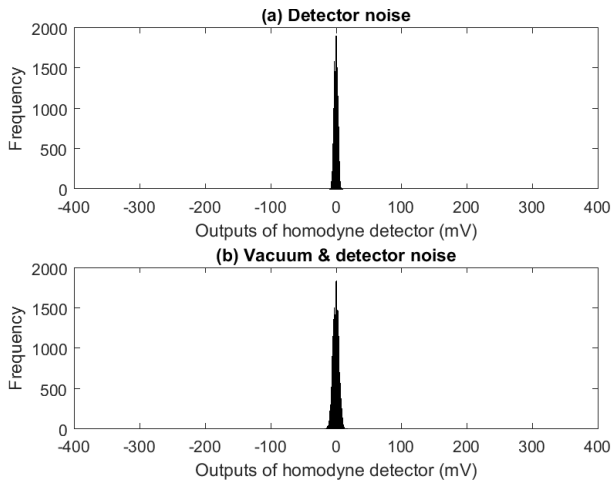}
	\captionsetup{justification=raggedright,
					singlelinecheck=false }
	\caption{Histograms of detector noise and vacuum noise. (a) Detector noise measured by turning off both the ASE source and the LO; (b) Vacuum (and detector) noise measured by turning on the LO only. Sample
size=32768.}
	\label{fig:3}
\end{figure}

In the second experiment, an 8-bit oscilloscope (Agilent) was used to acquire $10^7$ samples at a sampling rate of 100MHz. The autocorrelation of the collected data is shown in Fig.4. The correlation coefficients for Lags larger than zero are below $10^{-3}$ and within the range of the statistical uncertainty due to the finite sample size. This highlights the high quality of the entropy source.

\begin{figure}[t]
	\includegraphics[width=.5\textwidth]{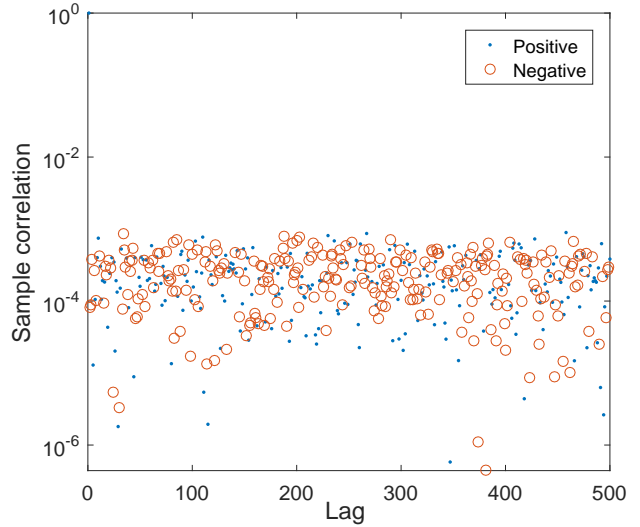}
	\captionsetup{justification=raggedright,
					singlelinecheck=false }
	\caption{Autocorrelation of raw data. Sample size=$10^7$.}
	\label{fig:4}
\end{figure}

While the above Gaussian-distributed raw data could be useful in certain applications, uniformly distributed random bits are more common in practice. There are different ways to generate binary random numbers from the Gaussian output of the homodyne detector. One simple way is to feed the Gaussian raw data into a voltage comparator, which is essentially a 1-bit ADC. By balancing the efficiencies of the two photodiodes and choosing a suitable threshold voltage, one random bit can be generated from each raw sample. To improve the random number generation rate, multi-bit ADC can be employed.

In our experiment, we use the internal 8-bit ADC of the oscilloscope to sample the output of the homodyne detector. An 8-bit ADC converts an analog input into one of the 256 ($2^8$) output bins. If the widths of the 256 bins have been tailored in a way so that the Gaussian distributed analog input will result a digital output uniformly distributed among the 256 bins, then we could generate 8 random bits per detection. However, since the ADC in our experiment has equal bin size, the digital outputs (raw samples) are not uniformly distributed. Here, we implement a Toeplitz hashing extractor \cite{WC81} to generate binary random bits from the 8-bit raw samples.

\begin{table*}[t]
\centering
\caption{The NIST statistical test results of $10^9$ random bits} \label{tab:1}

\begin{tabular}{|c|c|c|c|c|c|c|c|c|c|c|c|c|}
\hline 
\multicolumn{13}{|c|}{Results for the uniformity of P-values and the proportion of passing sequences} \\ 
\hline 
C1 & C2 & C3 & C4 & C5 & C6 & C7 & C8 & C9 & C10 & P-value & Proportion & Statistical test \\ 
\hline 
107 & 99 & 84 & 102 & 89 & 100 & 99 & 116 & 100 & 104 & 0.632955 & 992/1000 & Frequency \\ 
\hline 
102 & 78 & 85 & 106 & 94 & 108 & 93 & 113 & 100 & 121 & 0.088762 & 993/1000 & BlockFrequency \\ 
\hline 
110 & 101 & 91 & 99 & 98 & 89 & 96 & 95 & 110 & 111 & 0.769527 & 994/1000 & CumulativeSums \\ 
\hline 
98 & 91 & 98 & 98 & 94 & 115 & 91 & 116 & 104 & 95 & 0.603841 & 991/1000 & Runs \\ 
\hline 
96 & 104 & 95 & 115 & 112 & 101 & 80 & 111 & 93 & 93 & 0.314544 & 990/1000 & LongestRun \\ 
\hline 
111 & 86 & 77 & 107 & 123 & 88 & 100 & 108 & 97 & 103 & 0.057146 & 991/1000 & Rank \\ 
\hline 
93 & 102 & 118 & 96 & 105 & 95 & 101 & 103 & 94 & 93 & 0.800005 & 995/1000 & FFT \\ 
\hline 
104 & 93 & 96 & 91 & 102 & 102 & 95 & 107 & 99 & 111 & 0.932333 & 982/1000 & NonOverlappingTemplate \\ 
\hline 
98 & 105 & 97 & 108 & 94 & 86 & 100 & 92 & 119 & 101 & 0.574903 & 990/1000 & OverlappingTemplate \\ 
\hline 
94 & 100 & 112 & 94 & 107 & 93 & 102 & 101 & 101 & 96 & 0.948298 & 989/1000 & Universal \\ 
\hline 
95 & 97 & 114 & 81 & 101 & 98 & 112 & 97 & 111 & 94 & 0.431754 & 995/1000 & ApproximateEntropy \\ 
\hline 
67 & 51 & 62 & 69 & 76 & 63 & 63 & 62 & 64 & 53 & 0.592591 & 619/630 & RandomExcursions \\ 
\hline 
53 & 46 & 65 & 63 & 52 & 72 & 67 & 66 & 57 & 89 & 0.012031 & 621/630 & RandomExcursionsVariant \\ 
\hline 
117 & 90 & 118 & 115 & 75 & 101 & 91 & 93 & 104 & 96 & 0.044797 & 991/1000 & Serial \\ 
\hline 
97 & 92 & 91 & 88 & 104 & 106 & 120 & 88 & 96 & 118 & 0.194813 & 986/1000 & LinearComplexity \\ 
\hline 
\multicolumn{13}{|p{14cm}|}{
(1) The significant level $\alpha=0.01$. The first 10 columns in the table represent the distribution of P-values. The P-value column shows the uniformity of P-values, which should be larger than 0.0001 to pass the test.

(2) The minimum pass rate for each statistical test with the exception of the random excursion (variant) test is approximately = 980 for a sample size = 1000 binary sequences.

(3) The minimum pass rate for the random excursion (variant) test is approximately = 616 for a sample size = 630 binary sequences.

(4) For tests with multiple test results, the worst cases are presented.
} \\ 
\hline 
\end{tabular} 

\end{table*}

We represent the 8-bit output of the ADC as random variable $X$. The maximum number of random bits that can be extracted from each 8-bit raw sample is lower bounded by the min-entropy of $X$, which is defined as
\bes\label{eq4} H_{min}=-\log_2(P_{max}),\ees 
where $P_{max} = max_{x\in\lbrace 0,1\rbrace ^8} Pr[X = x]$ quantifies the maximal probability that $X$ to be one of the $2^8$ binary sequences. Using (4), the mini-entropy of the raw data has been determined to be $H_{min}=6.4$. This suggests that in the asymptotic case, we can generate on average 6.4 random bits from each 8-bit raw sample, corresponding to an efficiency of 0.8 bits per raw bit. We remark that if the untrusted technical noise of the system makes a significant contribution to the output, conditional min-entropy should be employed to estimate the extractable randomness \cite{HA15}.  

We apply a Toeplitz-hashing extractor to generate nearly perfect random bits from the raw data. A Toeplitz-hashing extractor extracts a m-bit random sequence by multiplying a n-bit raw sequence with a n-by-m Toeplitz matrix. The Toeplitz matrix is constructed from a long ($n+m-1$ bits) but reusable random seed. In our implementation, we choose m = 256 and n = 400, corresponding to an efficiency of 0.64 bits per raw bit (or 5.12 bits per raw sample). According to the leftover hash lemma \cite{IL89}, the security parameter $\varepsilon$ of the randomness extractor is determined by
\bes\label{eq5} m=n\frac{H_{min}}{8}-2\log_2(\dfrac{1}{\varepsilon}).\ees 
Using $n=400$, $m=256$, and $H_{min}=6.4$, the security parameter can be determined to be $\varepsilon=2^{-32}$, which means the statistical distance between the extracted random sequence and the prefect random sequence is bounded by $\varepsilon=2^{-32}$ \cite{ZN16}. To evaluate the quality of the random bits generated from the randomness extractor, 1 gigabits random numbers are generated and fed into the standard NIST statistical test suite \cite{NIST}. As summarized in Table 1, our random number generator passes all the 15 NIST tests. 

Limited by the bandwidth of the detector, the sampling rate used in our experiment is 100 Mega samples per second. Combined with the hashing efficiency of 5.12 bits per raw sample, the equivalent random number generation rate is 512 Mbps. To further increase the rate, one straightforward way is to use a detector with a larger bandwidth. For example, using a 15GHz detector, the sampling rate can reach 10 Giga samples per second (GSps) \cite{NH15}. The corresponding random number generation rate could be 51.2 Gbps. Of course, in practice, a faster detector tends to have a higher electrical noise which may reduce the quantum noise-to-electrical noise ratio. The achievable rate has to be evaluated based on actual characteristics of the system. One interesting question is given noiseless detector and ADC with arbitrarily high precision, what is the ultimate limit of the random number generation rate? This question has been addressed in \cite{MX13, HA15}. In fact, if the detector is noiseless and can resolve the photon number of the input state, then instead of performing homodyne detection, we can measure the photon number of the thermal state directly. The photon number distribution of a thermal state is given by \cite{L00}
\bes\label{eq6} P(n)=\frac{\langle n\rangle^n}{(1+\langle n\rangle)^{1+n}},\ees 
where $\langle n\rangle$ is the average photon number per sampling time window. It is easy to see the maximum probability is $P_{max}=P(0)=1/(1+\langle n\rangle)$. From (4), the min-entropy of the source is $H_{min}=log_2(1+\langle n\rangle)$. For example, if the output power of the thermal source is 1mw (at 1550nm) and the sampling rate is 10 GSps, then the average photon number within the sampling window (100ps) is about $7.8\times 10^5$. In principle, we could generate 19.6 random bits per sample.

One appealing feature of our scheme is its simple and robust design: both the incoherent source and the LO are operated in the CW mode; no active modulation or phase (or polarization) stabilization is required. Its compact design also opens the door to chip-size integration. In fact, a fully integrated QRNG on an InP platform has been demonstrated recently, where the laser sources, beam splitter and photodetectors are all integrated on a single chip \cite{AA16}. Since our scheme can be implemented with similar components, we expect the integration technologies developed there can also be applied in our scheme.

In summary, we propose and demonstrate a high quality photonic entropy source for random number generation. By mixing the output of a broadband incoherent source with a single mode LO at a beam splitter and performing differential photo-detection, the quadrature fluctuations of a single mode thermal state can be explored to generate true randomness. Experimentally, a quadrature variance about three orders of magnitude larger than the vacuum noise is observed, suggesting this scheme can tolerate much higher technical noises in comparison with QRNG based on measuring the vacuum noise. The high quality of this entropy source is evidenced by the small correlation coefficients of the acquired data. By applying Toeplitz hashing extractor on the Gaussian distributed raw data, unbiased random bits have been generated with an efficiency of 5.12 bits per sample. The output of the Toeplitz extractor successfully passes all the NIST statistical tests.

This work was performed at Oak Ridge National Laboratory (ORNL), operated by UT-Battelle for the U.S.
Department of Energy under Contract No. DE-AC05-00OR22725. The author acknowledges support from ORNL laboratory directed research and development program (LDRD), the U.S. Department of Energy Cybersecurity for Energy Delivery Systems (CEDS) program program under contract M614000329, and the U.S. Office of Naval Research (ONR).

\end{document}